\documentclass[aps,twocolumn,prb,showpacs,superscriptaddress,groupaddress]{revtex4}
\usepackage{amssymb}

\usepackage{amsmath}
\usepackage[mathcal]{eucal}
\usepackage{graphicx}
\usepackage{dcolumn}
\usepackage{amsfonts}
\usepackage{bm}
\usepackage{array}
\usepackage{natbib}

\newcommand{\be}{\begin{equation}}
\newcommand{\ee}{\end{equation}}
\newcommand{\bea}{\begin{eqnarray}}
\newcommand{\eea}{\end{eqnarray}}

\renewcommand{\vec}[1]{{\bm #1}}

\newcommand{\Tr}{\mathrm{Tr}}

\bibpunct{[}{]}{,\!}{n}{,}{,} 
\begin{document}

\title{Effects of interaction on field induced resonances in confined Fermi liquid}
\author{A. Iqbal}
\affiliation{Department of Physics and Astronomy, University of Iowa, Iowa City, Iowa 52242, USA}
\author{M. Khodas}
\affiliation{Department of Physics and Astronomy, University of Iowa, Iowa City, Iowa 52242, USA}
\begin{abstract}
We consider the two-dimensional electron gas confined laterally to a narrow channel by a harmonic potential.
As the Zeeman splitting matches the intersubband separation the nonlocal spin polarization develops a minimum as reported by Frolov {\it et al.} [Nature (London) {\bf 458}, 868 (2009)].
This phenomenon termed Ballistic Spin Resonance is due to the degeneracy between the nearest oppositely polarized subbands that is  lifted by spin-orbit coupling.
We showed that the resonance survives the weak and short-range interaction. 
The latter detunes it and as a result shifts the Zeeman splitting at which the minimum in spin polarization occurs.
Here this shift is attributed to the absence of Kohn theorem for the spin sloshing collective mode.
We characterized the shift due to weak interaction qualitatively by analyzing the spin sloshing mode within the Fermi liquid phenomenology.
\end{abstract}
\maketitle
\section{Introduction}

The manipulation of spins in non-magnetic structures by all electrical means is central to the technology of semiconductor spintronic devices \cite{Wolf2001,Awschalom2002,Hall2006,Awschalom2007}.
The spin relaxation processes cause the loss of information encoded in the spin degrees of freedom and represent a challenge for semiconductor technologies.

The geometrical confinement was argued to enhance spin correlations \cite{Bournel1998,Malshukov2000,Kiselev2000,Pareek2002}.
In the limiting case of a one-dimensional wire the spin relaxation is suppressed \cite{Schwab2006}. 
Experimentally, the increase in the spin relaxation length was observed in wires with thickness exceeding the mean free path \cite{Holleitner2006} which was attributed to the long lived homogeneous and spin spiral configurations \cite{Schwab2006,Chang2009,Froltsov2001,Pershin2005}.
Likewise, the in-plane magnetic field was argued theoretically \cite{Froltsov2001} and shown experimentally \cite{Meijer2004} to suppress the spin relaxation.

While separately, both the confinement and the magnetic field increase  the spin relaxation length,
their combination was experimentally shown to produce an opposite effect under certain conditions \cite{Frolov2009}.
In the experiment \cite{Frolov2009} spins were injected into a quasi-one-dimensional channel and detected via a spin-polarized quantum point contact.
The spin accumulation in the channel was non-monotonic function of the in-plane magnetic field $\vec{B}$ oriented perpendicular to the channel.
It reached a minimum for special values of the Zeeman splitting $E_Z$.
Quasi-classically, the dip in spin polarization was attributed to the commensuration between the spin precession time $2 \pi E_Z^{-1}$ (we henceforth set $\hbar = 1$) with twice the time it takes electrons to cross the channel in the direction of confinement, \cite{Luscher2010,Hachiya2014}.
Hence the phenomenon was termed the Ballistic Spin Resonance (BSR).
For the parabolic confinement as is assumed in this paper this time interval is the classical period of oscillations perpendicular to the channel $2 \pi \omega_c^{-1}$, where $\omega_c$ is the frequency of such oscillations.  
The condition for the BSR as stated in the Ref.~\cite{Luscher2010,Hachiya2014} can therefore be formulated as 
$E_Z^{-1} = \omega_c^{-1}$.

\begin{figure}[h]
\begin{center}
\includegraphics[width=1.0\columnwidth]{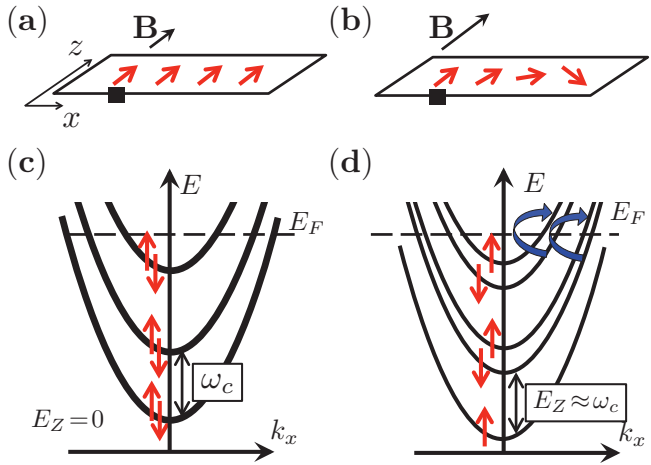}
\caption{ (color online) 
(a) The spins shown as (red) straight arrows injected into the channel via the quantum point contact denoted by a (black) square are polarized along the applied magnetic field, $\vec{B} \parallel \hat{z}$. 
The channel is running parallel to $x$ axis. 
Normally, they maintain their polarization in the course of propagation along the channel.
(b) At the BSR $E_Z \approx \omega_c$ the spins are still injected with the initial polarization along $\vec{B}$.
Their state of polarization however is modified as they move along the channel.
(c) In the absence of magnetic field the energy levels form subbands due to the quantization along $\hat{z}$.
For the parabolic confinement the subband splitting, $\omega_c$ is equal to the angular frequency of oscillations across the channel.
(d) At the BSR, $E_Z \approx \omega_c$ the adjacent subbands with opposite spin polarization are nearly degenerate.
The spin flip intersubband p-h excitations shown by thick round (blue) arrows become soft at the BSR, and play a central role in our analysis.}
\label{fig:setup}
\end{center}
\end{figure}

Recently one of us constructed the quantum mechanical description of the BSR \cite{Berman2014}.
The mechanism of the BSR according to the Ref.~\cite{Berman2014} is summarized as follows. 
Electrons injected via the quantum point contact are polarized parallel to the external field $\vec{B}$.
Normally, electrons maintain their polarization during their propagation along the channel, see Fig.~\ref{fig:setup}(a).
The special situation arises when the pairs of oppositely polarized states at nearest subbands are tuned into degeneracy as shown in Fig.~\ref{fig:setup}(d).
The magnetic field required for this satisfies 
$E_Z \approx \omega_c$ which is the quantum mechanical version of the semiclassical condition formulated in Ref.~\cite{Frolov2009}.
In this regime the spin-orbit interaction mixes the degenerate pairs of states even if nominally weak. 
The subband reconstruction taking place at $E_Z \approx \omega_c$  modifies drastically the spin dynamics inside the channel.
The injected spins no longer maintain their initial polarization, see Fig.~\ref{fig:setup}(b).
Instead, they oscillate between the original states defined in the absence of spin-orbit interaction.
These oscillations eventually lead to the decay of the total spin polarization down the channel.

In the ballistic regime the spin oscillations are washed out because of the finite spread in oscillation frequencies across subbands and/or due to the 
differences in arrival time from injector to detector.
Even though the BSR is robust in the presence of relatively strong short range disorder \cite{Berman2014} its actual shape is sensitive to the details of intersubband scattering \cite{Hachiya2014}.

In this work we analyze the effect of the electron-electron interaction on the BSR.
We start with the brief summary of our {\it main results}.

We consider the Fermi gas confined by a parabolic potential characterized by a frequency $\omega_c$ of the transversal motion. 
The local electron-electron interaction 
\begin{align}\label{M19}
\hat{H}_{\mathrm{int}} = \frac{1}{2} \int d \vec{r} \left[ V_{\rho} \left[\hat{\rho}(\vec{r})\right]^2 + 
V_s \hat{\vec{\sigma}}(\vec{r}) \cdot  \hat{\vec{\sigma}}(\vec{r}) \right]\,
\end{align}
is parametrized by amplitudes $V_{\rho}$ and $V_s$ of the interaction in the charge, $\hat{\rho}(\vec{r})$ and spin, $\hat{\sigma}(\vec{r})$ channels respectively.
Our main finding is that whereas for non-interacting system the BSR occurs at $B=B_{\mathrm{BSR}}$ such that the Zeeman splitting $E_Z = \omega_c$, in the presence of interaction Eq.~\eqref{M19} this field is renormalized,
\begin{align}\label{BSR_ren}
B^*_{\mathrm{BSR}} \approx B_{\mathrm{BSR}} \left[ 1 - 3 (V_{\rho} - 3 V_s) \nu/4 \right]\, .
\end{align}
The result \eqref{BSR_ren} holds for weak interaction, $\{ V_{\rho}\nu,  V_{s}\nu \} \lesssim 1$, where $\nu = m/\pi$ is the density of states of a Fermi gas of electrons of a mass $m$.
The result \eqref{BSR_ren} is obtained by combination of the two approaches: perturbation theory and the Fermi liquid phenomenology with local quasi-particle interaction.
The latter is applicable provided the channel width is much larger than the 
typical Fermi wavelength.
This condition is equivalent to the number of occupied subbands being large.
In the experiment \cite{Frolov2009} the Fermi energy of the electron gas at was $E_F \approx 4$meV corresponding to the density of $\approx 1.1 \times 10^{11}$cm$^{-2}$ in the high mobility GaAs-based heterostructures.
The reported magnetic field at the onset of the BSR on the other hand was $\approx 7$T. 
This gives $\omega_c \approx 0.15$meV for the bulk $g$-factor of $-0.44$.
The number of occupied subbands was therefore $E_F/\omega_c \approx 27 \gg 1$, and the Fermi liquid description applies.

In a Fermi gas when the Zeeman splitting is tuned into the resonance, $E_Z = \omega_c$ the spin flip intersubband  particle-hole (p-h) excitations shown by round arrows in Fig.~\ref{fig:setup}(d) become soft, i.e. with nearly zero excitation energy.
Perturbation theory shows that although the energy of each such pair is modified by interaction, different p-h remain degenerate. 
It follows that the BSR can be meaningfully defined for $\{ V_{\rho}\nu,  V_{s}\nu \} \lesssim 1$.
As interaction causes the detuning of the BSR the condition for the BSR is modified in turn.

The second ingredient leading to Eq.~\eqref{BSR_ren} is the relation of the BSR to the spin density oscillations across the channel, or to the collective spin sloshing mode (see Fig.~\ref{fig:slosh}).
The perturbation theory indicates that the energy of this mode is very close to the spin flip intersubband  p-h excitations.
This observation links the BSR to the collective behavior.
We studied the spin sloshing mode within the phenomenological Fermi and obtained for its frequency
\begin{align}\label{result1_b}
\omega_s(E_Z) \approx 
E_Z  - \omega_c +(V_{\rho} \! -\!  3 V_s)\nu
(\omega_c/2 + E_Z/4)
\end{align}
valid under the same conditions as Eq.~\eqref{BSR_ren}.
The latter readily follows from Eq.~\eqref{result1_b} since the BSR sets in for $E_Z = E_Z^*$ such that $\omega(E_Z^*)  = 0$.

As evidenced by the result \eqref{result1_b} the $B_{\mathrm{BSR}}$ is  renormalized because of the absence of what would be the equivalent of Kohn theorem \cite{Kohn1961} for spin sloshing mode.
At first glance the Kohn theorem for spin sloshing mode should hold.
Indeed, in the parabolic potential the frequency of the density oscillations across the channel, or of the so called sloshing mode is $\omega_c$ regardless of interaction.
This was first discussed theoretically  \cite{Dobson1994,Brey1989} and confirmed experimentally \cite{Wixforth1994} in the context of semiconductor quantum wells, and more recently in relation to trapped cold fermions in optical lattices, \cite{Minguzzi2001,Chiacchiera2009}. 
Similarly, the Zeeman splitting gives rise to a collective spin precession mode which is similarly not renormalized according to another version of the Kohn theorem \cite{Yafet1963}.
As neither $\omega_c$ nor $E_Z$ is renormalized it is tempting to conclude that the frequency of the spin sloshing mode assumes its
non-renormalized value, $\omega_s = | \omega_c - E_Z|$.
Correspondingly, the spin sloshing mode and the p-h excitations degenerate with it soften at $E_Z = \omega_c$.
As a result one would mistakenly conclude that interactions do not modify the condition for BSR. 

To see why the BSR {\it is} nevertheless shifted by the interaction in accordance with Eq.~\eqref{BSR_ren} it is necessary to identify the collective degree of freedom unaffected by interactions whenever the Kohn theorem applies. 
In the case of the sloshing mode such degree of freedom is the center of mass, $d_z =\sum_i z_i$.
Here $z_i$ is the location of the $i$th electron.
In the case of spin precession such observable is the total spin, $\vec{\sigma} = \sum_{i} \vec{\sigma}_i$. 
The collective variable associated with the spin sloshing mode is $\sum_{i} z_i \vec{\sigma}_i$ inducing the transitions shown in Fig.~\ref{fig:setup}(d).
In contrast to $d_z$ and $\vec{\sigma}$, the observable $\sum_{i} z_i \vec{\sigma}_i$ in general does not commute with the interaction Hamiltonian.
This is the underlying reason for the absence of the Kohn theorem for the spin sloshing mode.
The situation similar in this regard arises for the chiral spin resonance when the transitions between spin-orbit split bands are induced by an ac electric field \cite{Shekhter2005}.
In the remainder of the paper we expand upon the above results.

The paper is organized as follows.
In Sec.\ref{sec:Fermi} we develop the hydrodynamic approach based on Fermi liquid to describe the three collective excitations: sloshing mode, spin precession mode and spin sloshing mode analyzed in Secs.~\ref{sec:Sloshing_mode}, \ref{sec:Collective_spin} and \ref{sec:Collective_spin-sloshing} respectively.
In Sec.~\ref{sec:Microscopic} we analyze on the spin-sloshing mode using the perturbation theory.
We compare the results obtained in Secs.~\ref{sec:Collective_spin-sloshing} and \ref{sec:Microscopic} in Sec.~\ref{sec:Comparison}.
Following the comparison the effect of interaction on BSR is presented in Sec.~\ref{sec:Interaction}.
The results are summarized and discussed in Sec.~\ref{sec:Conclusions}. 
Some of the technicalities are relegated to appendices.

\section{Fermi liquid theory of spin sloshing mode}
\label{sec:Fermi}

In this section we study the three types of collective excitations in parabolically confined Fermi liquid: the sloshing mode, the spin precession and the spin sloshing mode.
Crucially, while the BSR requires finite spin-orbit interaction, the condition for its onset does not depend on it.
For that reason here we limit the consideration disregarding the spin-orbit interaction.
See Ref.~\cite{Chen1999,Ashrafi2012,Ashrafi2013} for the discussion of the Fermi liquid effects specific for systems with spin-orbit interaction.

\subsection{Sloshing mode in Fermi liquid confined by a parabolic potential} 
\label{sec:Sloshing_mode}

The phenomenological Fermi liquid is formulated in terms of the weakly interacting quasi-particles. 
Introduce the distribution function $n_{\vec{p}}(\vec{r})$  is the density of quasi-particles at the point of the phase space specified  by the coordinate $\vec{r}$ and momentum $\vec{p}$.
The function $n_{\vec{p}}(\vec{r})$ satisfies the transport equation, \cite{Pines1966} 
\be\label{FL1}
\partial_t n_{\vec{p}}(\vec{r})  + 
\{ \tilde{\epsilon}_{\vec{p}}(\vec{r}), n_{\vec{p}}(\vec{r}) \} = I[ n_{\vec{p}}(\vec{r})], 
\ee
where the Poisson bracket of the two functions $A_{\vec{p}}(\vec{r})$
and $B_{\vec{p}}(\vec{r})$ is defined by $\{ A, B\} = \partial_{\vec{p}} A \partial_{\vec{r}} B -
\partial_{\vec{r}} A \partial_{\vec{p}} B$.
The effective Hamiltonian  in Eq.~\eqref{FL1} 
\be\label{FL3}
\tilde{\epsilon}_{\vec{p}}(\vec{r}) = \epsilon_{\vec{p}}(\vec{r}) + \sum_{\vec{p}'} f_{\vec{p}\vec{p}'} \delta n_{\vec{p}'}
\ee
contains the quasi-particle energy $\epsilon_{\vec{p}}(\vec{r})$ and the energy of the interaction with the deformation of the Fermi surface described by the non-equilibrium part of the distribution function,
$\delta n_{\vec{p}}$ \cite{Pines1966}.
The coefficient $f_{\vec{p}\vec{p}'}$ in Eq.~\eqref{FL3} is the second variational derivative of the free energy with respect to $\delta n_{\vec{p}}$.
In this subsection we ignore the spin for clarity.
Finally, the right hand side of the Eq.~\eqref{FL3} is the collision term controlling the relaxation processes.

Let the confining potential be $U(z) = k z^2/2$ and the equilibrium distribution function $n_0( \epsilon_{\vec{p}}(z) )$.
The Kohn mode is the solution of Eq.~\eqref{FL1} of the form \cite{Chiacchiera2009,Pantel2012},
\begin{align}\label{FL5}
\delta n^K_{\vec{p}}(z) = 
[\partial_{ \epsilon_{\vec{p}}} n_0( \epsilon_{\vec{p}}(z) )]e^{- i \omega t}
\left( a_{\omega} z   + b_{\omega} \cos \theta_{\vec{p}} \right)
\end{align}
describing oscillations of the particle and current density across the channel at frequency $\omega$.
In Eq.~\eqref{FL5} we denote by $\theta_{\vec{p}}$ the angle the quasi-particle momentum forms with $z$-axis directed perpendicular to the channel, see Fig.~\ref{fig:setup}(a).
It follows that $\cos \theta_{\vec{p}} = p_z/ p_F$, where $p_F(z)$ is the Fermi momentum.
The density oscillations represented by Eq.~\eqref{FL5} have a vanishing amplitude at the center of the channel, $z=0$, and grow linearly away from it.
The equation \eqref{FL5} therefore describes the collective sloshing of the Fermi liquid across the channel, and is referred to as the sloshing mode.

We show that in the Galilean invariant system the non-trivial solutions to Eq.~\eqref{FL1} of the form \eqref{FL5} exists for the choice $\omega = \omega_c= \sqrt{k/m}$ determined by the bare electron mass and not sensitive to the interaction.
The statement of the existence of the sloshing mode with unrenormalized frequency $\omega_c$  is general and does not depend on the statistics or temperature. 
Below we demonstrate it for the Galilean invariant Fermi liquid, setting the stage for the discussion of the spin sloshing mode.

Thanks to the particle and momentum conservation, $I[\delta n^K]=0$ and substitution of Eq.~\eqref{FL3} into Eq.~\eqref{FL1} yields to the linear order in $\delta n^K$,
\begin{align}\label{FL6}
- i \omega \delta n^K + \{\sum_{\vec{p'}} f_{\vec{p}\vec{p}'} \delta n^K_{\vec{p}'}, n^0 \} 
+
\{ \epsilon_{\vec{p}},  \delta n^K \} = 0\, .
\end{align}
In the second term of Eq.~\eqref{FL6} we have 
\begin{align}\label{FL7}
\sum_{\vec{p}'} f_{\vec{p}\vec{p}'} \delta n^K_{\vec{p}'} = 
-e^{ -i \omega t} \left[F_0 z a_{\omega} + F_1 b_{\omega} \cos \theta_{\vec{p}}\right]\, ,
\end{align}
where the Landau parameters are defined by the Fourier expansion 
$\nu^* f_{\vec{p}\vec{p}'} = \sum_{m=-\infty}^{\infty} F_m  e^{i m (\theta_{\vec{p}}-\theta_{\vec{p}'})}$.
Here quasi-particle density of states $\nu^* = m^*/\pi$ is proportional to the effective mass $m^*$.

\begin{figure}[ht!]
\begin{center}
\includegraphics[width=1.0\columnwidth]{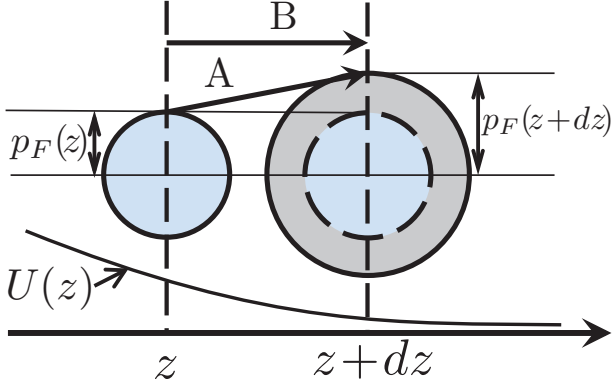}
\caption{ (color online) 
The two Fermi surfaces located at $z$ and $z+dz$ shown as solid circles.
As the confining potential $U(z)$ decreases with $z$ the Fermi momenta at the two points satisfy $p_F(z+dz) > p_F(z)$.
When a quasi-particle is transferred from $z$ to $z+dz$ its energy changes partly due to the interaction with the quasi-particles in the annulus $p_F(z) <  p < p_F(z+dz)$ (grey) of a radius $dp_F$. 
This interaction is an ultimate cause of the force renormalization, Eq.~\eqref{FL11}.
For the quasi-particle transferred along the path A the energy change is zero.
The translation along path B with momentum kept constant is used to derive Eq.~\eqref{FL11}.}
\label{fig:FL1}
\end{center}
\end{figure}

The momentum derivative of the equilibrium distribution function in Eq.~\eqref{FL6} reads
\begin{align}\label{FL9}
\partial_{p_z} n_0 =\left[ \partial_{\epsilon} n_0\right]\left( p_z/ m^*\right)
\end{align}
as by definition $1/m^* = \partial_{p_z} \epsilon_{\vec{p}}(z) / p$ evaluated at $p = p_F$.
For the spatial derivative we have, $\partial_z n_0= \left[\partial_{\epsilon} n_0\right] \partial_z \epsilon_{\vec{p}}(z)$, where
\begin{align}\label{FL11}
\partial_z \epsilon_{\vec{p}}(z)=  \frac{ \partial_z U(z) }{ 1 + F_0 }\, .
\end{align}
The equation \eqref{FL11} can be understood as follows.
Let the quasi-particle be translated from the Fermi surface at $z$ to the Fermi surface at $z+dz$ as shown in Fig.~\ref{fig:FL1} as path A.
At equilibrium the free energy should be the same before and after the translation,
\begin{align}\label{FL13}
 U(z\!+\! d z) - U(z) \!=\!v_F\! \left[ p_F(z) - p_F(z\!+\!dz) \right]\!(1 + F_0),
\end{align}
where $v_F = p_F/ m^*$ is the Fermi velocity.
The term $\propto F_0$ accounts for the interaction of the trial quasi-particle with the Fermi surface deformation it sees as a result of a spatial translation by $dz$ see Fig.~\ref{fig:FL1}.
Equation \eqref{FL13} gives 
\begin{align}\label{FL15}
v_F \frac{ d p_F(z) }{ d z } = - \frac{ \partial_z U}{ 1 + F_0}\, .
\end{align}
We now translate the quasi-particle from $z$ to $z+dz$ while keeping its momentum constant, as shown in Fig.~\ref{fig:FL1} as path B.
The energy change as a result of this translation is
\begin{align}\label{FL17}
d \epsilon_{\vec{p}}(\vec{r}) = \left[U(z+dz) - U(z)\right] + v_F d p_F F_0\, , 
\end{align}
where the first term is the change in the confining potential, and the second term is due to the interaction with an extra quasi-particles with momenta in the annulus of the width $dp_F$, $p_F(z)< p < p_F(z+dz)$.
Combination of Eqs.~\eqref{FL15} and \eqref{FL17} gives Eq.~\eqref{FL11}.

Substituting Eqs.~\eqref{FL7}, \eqref{FL9} and \eqref{FL11} in Eq.~\eqref{FL6}, and equating both the linear in $z$ and linear in $p_z$ parts to zero we obtain,
\begin{align}\label{FL21}
i \omega a_{\omega} - b_{\omega} \frac{ k}{ p_F}\frac{ 1 + F_1 }{ 1 + F_0}  &= 0
\notag \\
a_{\omega} v_F ( 1 + F_0) + i \omega b_{\omega}   & = 0 \, .
\end{align}
The non-trivial solution of \eqref{FL21} is obtained for $\omega^2 = (k/m^*)(1 + F_1) $.
In view of the Galilean invariance
\begin{align}\label{Galileo}
m^*/m = 1 + F_1\, ,
\end{align}
and we recover the Kohn mode with frequency $\omega = \omega_c$.
%
%
\subsection{Collective spin precession mode in a confined Fermi liquid}
\label{sec:Collective_spin}
We now consider the collective spin precession mode in the presence of an in-plane magnetic field $\vec{B}$.
For definiteness we assume that $\vec{B} \parallel \hat{z}$. 
In the absence of interactions spin of $i$-th electron, $\vec{\sigma}_i$ precesses with the frequency $E_Z$ equal to the Zeeman splitting.
As the spin conserving interaction commutes with the total spin, 
the latter precesses at the same unrenormalized frequency $E_Z$ as in the non-interacting case.

We show that this statement holds when the Fermi liquid is spatially confined.
In order to describe the spin dynamics we consider the distribution function as a density matrix in spin space, and generalize the transport equation \eqref{FL1} to \cite{Pitaevskii1980}
\begin{align}\label{kin_eq}
\partial_t  \hat{n}_{\vec{p}}(\vec{r}) \!+\! i[\hat{\tilde{\epsilon}}_{\vec{p}}(\vec{r}),\! \hat{n}_{\vec{p}}(\vec{r})]\! +\!
\{\hat{\tilde{\epsilon}}_{\vec{p}}(\vec{r}), \!\hat{n}_{\vec{p}}(\vec{r})\}\!=\! I[\hat{n}_{\vec{p}}(\vec{r})],
\end{align}
where the commutator of the two matrices is $[A,B] = AB - BA$.
The quasi-particle Hamiltonian 
\begin{align}\label{SP1}
\hat{\tilde{\epsilon}}_{\vec{p}}\! =\! 
\epsilon_{\vec{p}}\! +\! \sum_{\vec{p}'}\left[ g_{\vec{p}\vec{p}'} \vec{\sigma} \Tr (\vec{\sigma}' \hat{\delta n}_{\vec{p}'})
+ f_{\vec{p}\vec{p}'} \Tr ( \hat{\delta n}_{\vec{p}'})\right]
-\frac{E_Z^r}{2}\sigma_z\, ,
\end{align}
where Tr stands for the trace over spin indices.
The last term on the right hand side of Eq.~\eqref{SP1} is the renormalized Zeeman coupling, \cite{Pitaevskii1980} 
\begin{align}\label{Zeeman_bare}
E_Z^r = E_Z(1 +G_0)^{-1}\, . 
\end{align}
Here and below we neglect the dependence of the Fermi liquid parameters on the magnetic field.
Such an approximation is satisfied as long as $E_F \gg E_Z$.
In a particular setup of Ref.~\cite{Frolov2009} the above condition is safely satisfied.
The collective spin precession is described by the density matrix 
\begin{align}\label{SP2}
\hat{n}_{pr} = n_0  + [\partial_{\epsilon} n_0] \frac{E_Z^r}{2}\sigma_z + [\partial_{\epsilon} n_0] e^{-i \omega t} A \sigma_{\pm}\, ,
\end{align}
where $\sigma_{\pm} = \sigma_x \pm i \sigma_y$.
The second term of Eq.~\eqref{SP2} stands for the equilibrium polarization of the Fermi liquid.
Due to the spin conservation, $I[\hat{n}_{pr} ]  = 0$.
As the precession amplitude $A$ is position independent the Poisson bracket term of Eq.~\eqref{kin_eq} vanishes for the solution of the form \eqref{SP2}.
The quasi-particle energy Eq.~\eqref{SP1} takes the form
\begin{align}\label{SP3}
\hat{\tilde{\epsilon}}_{\vec{p}} = 
\epsilon_{\vec{p}} 
- \frac{E_Z^r}{2}\sigma_z
- G_0 A \sigma_{\pm} e^{-i \omega t} 
\end{align}
for the distribution function Eq.~\eqref{SP2}.
Here we have introduced $\nu^* g_{\vec{p}\vec{p}'} = \sum_{m=-\infty}^{\infty} G_m e^{i m (\theta_{\vec{p}}-\theta_{\vec{p}'})}$.
We obtain for the commutator of Eq.~\eqref{SP2} and Eq.~\eqref{SP3}
\begin{align}\label{SP4}
[\hat{\tilde{\epsilon}}_{\vec{p}},\hat{n}_{pr}] = -A E_Z^r (1 + G_0) \sigma_{\pm} e^{-i \omega t}\, .
\end{align}
Substitution of Eqs.~\eqref{SP4} in Eq.~\eqref{kin_eq} shows that the equation \eqref{SP2} is a solution with unrenormalized frequency $\omega = E_Z^r (1 + G_0) = E_Z$ as expected.
\subsection{Collective spin-sloshing mode}
\label{sec:Collective_spin-sloshing}
\begin{figure}[ht!]
\begin{center}
\includegraphics[width=0.9\columnwidth]{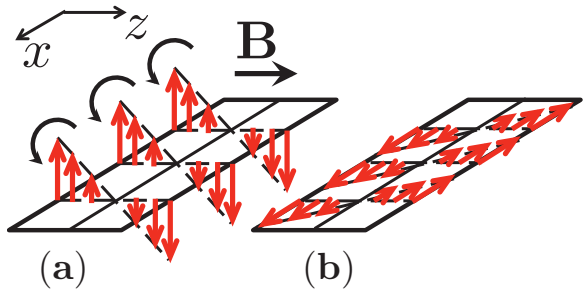}
\caption{ (color online) The representation of the spin sloshing mode.
The magnetic field $\vec{B}$ is assumed to point in $z$-direction perpendicular to the channel running along $x$-direction. 
Thick arrows (red) represents the spin polarization changing across the channel, and vanishing at its center, $z=0$.
The profile is translationally invariant along the channel and all the spins collectively precesses at frequency $\omega_s$.
(a) The snapshot of spin polarization at an instant such that the polarization is perpendicular to the $xz$-plane.
(b) In a quarter period of precession, $\pi/2 \omega_c$ the spin polarization starting in a configuration shown in (a) rotates by $90^{\circ}$ and becomes parallel to the $xz$-plane.}
\label{fig:slosh}
\end{center}
\end{figure}

In this section we construct and analyze the spin sloshing mode.
It combines features of both Kohn mode and collective spin precession mode discussed in Sec.~\ref{sec:Sloshing_mode} and \ref{sec:Collective_spin} respectively.
This mode is important because as we will show in Sec.~\ref{sec:Interaction} it provides us with the information on the interaction induced renormalization of the BSR.
In the spin sloshing mode the density does not change but the spin density undergoes precession with the amplitude vanishing at the center of the channel and growing linearly away from it, see Fig.~\ref{fig:slosh}.

Accordingly, the trial solution of the transport equation \eqref{kin_eq} describing the spin sloshing mode is 
\begin{align}\label{ansatz}
\hat{n}_{s} =  n_0   + [\partial_{\epsilon} n_0] 
\left(
e^{- i \omega t}( a_{\omega} z + b_{\omega} \cos \theta_{\vec{p}}) \sigma_{\pm} - \frac{E_Z^r}{2}\sigma_z  \right)\, ,
\end{align}
where similar to Sec.~\ref{sec:Collective_spin} we assume $\vec{B} \parallel \hat{z}$.
We look for the frequencies $\omega$ such that Eq.~\eqref{ansatz} is the non-trivial solution of Eq.~\eqref{kin_eq} with $a_{\omega}^2 + b_{\omega}^2 \neq 0$. 
The quasi-particle Hamiltonian, \eqref{SP1} for the density matrix \eqref{ansatz} reads
\begin{align}\label{slosh1}
\hat{\tilde{\epsilon}}_{\vec{p}} = 
\epsilon_{\vec{p}} - \frac{E_Z^r}{2}\sigma_z
- e^{-i \omega t} ( a_{\omega} z G_0 + b_{\omega} G_1\cos \theta_{\vec{p}}) \sigma_{\pm}  \, .
\end{align}
We now substitute \eqref{ansatz} and \eqref{slosh1} in the transport equation, Eq.~\eqref{kin_eq} and find the frequency $\omega$ at which the expression \eqref{ansatz} is a non-trivial solution.

Unlike the particle and spin density the spin-current in general is not conserved and the collision term $I[\hat{n}_s]$ is in general non-zero.
Here we limit the consideration to the collisionless regime.
On general grounds collisions lead the attenuation of the spin sloshing collective mode.
For the low temperatures studied in the experiment \cite{Frolov2009} electron-electron collision induced broadening can be assumed to be negligible.
 
In contrast to Eq.~\eqref{SP3} obtained for the collective spin precession Eq. \eqref{slosh1} contains the coordinate $z$ explicitly.
It follows that both the commutator and the Poisson brackets terms of Eq.~\eqref{kin_eq} are non-zero for Eq.~\eqref{ansatz}.
The commutator term in \eqref{kin_eq} can be easily computed yielding 
\begin{equation}\label{slosh2}
 [ \hat{\tilde{\epsilon}}, \hat{n}] =  
\mp  E_Z^r e^{-i \omega t} \sigma_{\pm}  [\partial_{\epsilon} n_0 ]\left( a_{\omega} z G_0^+ + b_{\omega} \cos \theta_{\vec{p}} G_1^+ \right),
\end{equation}
where we have introduced the notation $G_n^+ = G_n+1$ and similarly,
$F_n^+ = F_n +1$ for shortness.

The linearized expression for the Poisson bracket term in Eq.~\eqref{kin_eq} can be obtained from the following expressions for the derivatives of the quasi-particle Hamiltonian, Eq.~\eqref{slosh1}
\begin{subequations}\label{slosh3}
\begin{align}\label{slosh3a}
\partial_z \hat{\tilde{\epsilon}} & = \partial_z \epsilon_{\vec{p}} - e^{-i \omega t } a_{\omega} G_0 \sigma_{\pm}
\end{align}
\begin{align}\label{slosh3b}
\partial_{p_z} \hat{\tilde{\epsilon}} & = \frac{p_z}{m^*} - e^{-i \omega t } b_{\omega}  \frac{G_1}{p_F} \sigma_{\pm}\, ,
\end{align}
\end{subequations}
and the derivatives of the density matrix, Eq.~\eqref{ansatz}
\begin{subequations}\label{slosh4}
\begin{align}\label{slosh4a}
\partial_z \hat{n}_s & = [\partial_{\epsilon} n_0]
\left( \partial_z \epsilon_{\vec{p}} + a_{\omega} e^{-i \omega t} \sigma_{\pm} \right)
\end{align}
\begin{align}\label{slosh4b}
\partial_{p_z} \hat{n}_s & = [\partial_{\epsilon} n_0]
\left( \frac{p_z}{m^*} + \frac{b_{\omega}}{p_F} e^{ - i \omega t} \sigma_{\pm} \right)\, .
\end{align}
\end{subequations}
In writing Eq.~\eqref{slosh3a} the spatial derivatives of the Fermi liquid parameters were neglected.
These derivatives are expected to be small for large number of occupied subbands.
The reason is the slow variation of the Fermi liquid parameters in this high density and shallow confinement regime.
Note that in the considered case of a confined two-dimensional systems with weak and short-range interaction this approximation becomes exact because of the independence of the density of states on the particle concentration in two dimensions.
This justification is elaborated in Sec.~\ref{sec:Comparison} where we compare the results of Fermi liquid analysis with the microscopic calculations.

Using the Eqs.~\eqref{slosh3} and \eqref{slosh4} we obtain for the linearized Poisson bracket entering Eq.~\eqref{kin_eq}
\begin{align}\label{slosh5}
\{\hat{\tilde{\epsilon}}, \hat{n}_s\}=
[\partial_{\epsilon} n_0] e^{- i \omega t} \sigma_{\pm}
\left(
a_{\omega}G_0^+\frac{p_z}{m^*} - \frac{b_{\omega}}{p_F}G_1^+ \partial_z \epsilon_{\vec{p}}
\right).
\end{align}
Substitution of Eqs.~\eqref{slosh2} and \eqref{slosh5} along with Eq.~\eqref{FL11} and the expression for the time derivative $ \partial_t \hat{n} =- i \omega e^{-i \omega t} ( a_{\omega} z+ b_{\omega} \cos \theta_{\vec{p}}) \sigma_{\pm}$ into Eq.~\eqref{kin_eq} yields in the collisionless regime two conditions
\begin{align}\label{slosh6}
i  v_F a_{\omega}  \left( \omega \pm E_Z^r G_0^+\right) 
+b_{\omega}  \frac{k}{m^*} \frac{G_1^+}{F_0^+} &= 0
\notag \\
i  v_F a_{\omega}  G_0^+   + b_{\omega}  \left( \omega \pm E_Z^r G_1^+ \right) 
&=0\, .
\end{align}
The non-trivial solution of Eqs.~\eqref{slosh6} is obtained for $\omega = \omega_s$ and $\omega = \omega_f$ where
\begin{align}\label{result1}
\omega_{s,f} = 
& \left\vert 
 E_Z\frac{G_1^+  + G_0^+}{2  G_0^+}
\right.
\notag \\
&\,\, \left. \mp \sqrt{ \left[E_Z \frac{ (G_1^+ - G_0^+) }{  2 G_0^+ }\right]^2+ \omega_c^2 \frac{G_0^ +G_1^ +}{F_0^ +F_1^ +} }\right\vert \, .
\end{align}
The result Eq.~\eqref{result1} is stated in terms of bare frequencies $E_Z$ and $\omega_c$ through the Galilean invariance condition Eq.~\eqref{Galileo} and the relation \eqref{Zeeman_bare}.

The slow and fast collective modes with eigenfrequencies $\omega_s$ and $\omega_f$ are distinguished by their non-interacting limit.
In a free gas the two frequencies become $\omega_{f,s} = \left\vert\omega_c \pm E_Z\right|$.
Therefore, for not too strong interactions, at the onset of BSR the slow mode with frequency $\omega_s$ softens down. 
It is for this reason we henceforth focus on this mode.

We define $E_Z^*$ as the Zeeman splitting at which $\omega_s = 0$.
In the non-interacting Fermi gas we have $E_Z^* = \omega_c$.
In the Fermi liquid Eq.~\eqref{result1} gives
\begin{align}\label{result2}
E_Z^* = \omega_c \frac{ G_0^+ }{ \sqrt{ F_0^{+} F_1^{+}}}\, .
\end{align}
Both Eqs.~\eqref{result1} and \eqref{result2} refer to the collective excitations of a confined Fermi liquid.
The BSR on the other hand occurs when the individual quasi-particles satisfy the resonant condition.
Nevertheless, we demonstrate that the result \eqref{result2} bears on the renormalization of the BSR in the limit of weak interactions.
To this end in the next section we supplement the Fermi liquid phenomenology with the microscopic analysis.

\section{Microscopic analysis: Perturbation Theory} 
\label{sec:Microscopic}
Our goal is to compute the frequency of the collective mode, $\omega_s$  in perturbation theory in interaction.
The microscopic calculation will provide us with the independent way to confirm the results Eq.~\eqref{result1} and \eqref{result2} obtained phenomenologically.
More importantly, it will allow us to relate the interaction induced shift of the BSR to the renormalization of the collective mode as given by Eq.~\eqref{result1}.
Here we focus on the spin sloshing mode.
The analogous analysis of the sloshing and spin precession modes is detailed in App.~\ref{app:Kohn}.

The Hamiltonian 
\begin{align}\label{M11}
\hat{H} = \hat{H}_0 + \hat{H}_{\mathrm{int}}\, ,
\end{align}
where the second term given by Eq.~\eqref{M19} describes the interaction, and the first is a free quadratic part,
\begin{align}\label{M13}
\hat{H}_0 = \sum_{n,\alpha,k_x} \psi_{n,\alpha;k_x}^{\dag} E_{n\alpha}(k_x)  \psi_{n,\alpha;k_x}\, ,
\end{align}
where  $\psi^{\dag}_{n,\alpha;k}$ is the operator creating an electron in the state $|n,k_x\rangle | \alpha \rangle$ such that
\begin{align}\label{M14}
\langle z,x |n,k_x \rangle  = \varphi_n(z/\ell) e^{i k_x x}  \, .
\end{align}
The first factor of the wave function \eqref{M14} is the standard harmonic oscillator wave functions  
\begin{align}\label{M14a}
\varphi_n(\xi) = (2^m m!)^{-1/2} \pi^{-1/4} \exp( - \xi^2/2  ) H_n( \xi )\, ,
\end{align}
where $H_n(\xi)$ is the Hermite polynomials, and the subband index $n$ is non-negative integer.
The oscillator length is $\ell = 1/ \sqrt{ m \omega_c}$.
The exponential prefactor describes the propagation of the electron with the momentum $k_x$ along the channel.
The spinors $|\alpha\rangle$ satisfy $\sigma_z |\alpha\rangle = \alpha |\alpha\rangle$ and for $\alpha = \pm 1$ describe the state of polarization along the in-plane magnetic field $\vec{B} \parallel \hat{z}$.
The dispersion relation of non-interacting electrons is 
\begin{align}\label{M17}
E_{n\alpha}(k_x) = \frac{k_x^2}{2 m } + \omega_c \left( n +\frac{1}{2}\right) - E_Z\frac{\alpha}{2} \, .
\end{align}
For fixed $n$ and $\alpha$ the states for all possible $k_x$ form subbands labeled here by $|n,\alpha\rangle$ for shortness.
For a given Fermi energy, $E_F$ if $E_{n\alpha}(k_x=0)  < E_F$ the subband $|n,\alpha \rangle$ crosses the Fermi level at the Fermi momentum
\begin{align}\label{kF}
k^F_{n,\alpha} = \sqrt{2 m \left[E_F - \omega_c \left( n +\frac{1}{2}\right) + E_Z\frac{\alpha}{2} \right] }\, .
\end{align}
It will be convenient to set $k^F_{n,\alpha} = 0$ for bands not crossing the Fermi level.

\begin{figure}[h]
\begin{center}
\includegraphics[width=0.9\columnwidth]{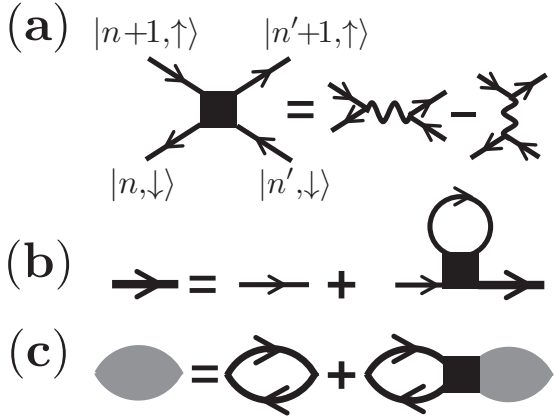}
\caption{(a) The scattering processes that leads to the most singular contributions to the correlation functions has both the initial and final states at resonance with the external frequency.
To the leading order this amplitudes contains the direct and exchange terms.
(b) The Dyson equation for the dressed Green function, Eq.~\eqref{Green} shown by the thick arrowed line combining the most singular contributions at the mass shell.
The thin line stands for the Green function of free electrons. 
(c) The Dyson equation for the matrix correlation function $\hat{\Gamma}$, Eq.~\eqref{Gamma} shown by the shaded area.
For the free electron gas $\hat{\Gamma}$ is given by the polarization operator, Eq.~\eqref{Polarization}.}
\label{fig:scattering}
\end{center}
\end{figure}

Our strategy is to sum exactly all the terms of perturbation theory in the interaction Eq.~\eqref{M19} of the form $\propto \left[ V_{\rho} (\omega - \delta \omega)^{-1} \right]^{n_1} \left[ V_{s} (\omega - \delta \omega)^{-1}\right]^{n_2} (\omega - \delta \omega)^{-1}$ with $n_{1,2}$ being a non-negative integer and we introduced the detuning 
\begin{align}\label{detuning}
\delta \omega = | E_Z - \omega_c |\, .
\end{align}
Such a procedure amounts to the summation of the most singular contributions for $\omega \approx \delta \omega$ at each order of perturbation theory.
It is accurate provided the interaction is sufficiently weak.
More precisely, the typical frequency shift due to the interaction should be smaller than the intersubband separation, 
$V_{\rho,s} \ell^{-2} \lesssim \omega_c$.
Introducing the unrenormalized density of states $\nu = m / \pi$ we arrive at the condition $\left\{ V_{\rho} \nu, V_s \nu \right\} \lesssim 1$.

To select the graphs giving the most singular contributions for $\omega \approx \delta \omega$ note that electron transitions between $|n+1,k_x\rangle |\alpha=+1 \rangle$ and $|n,k_x\rangle |\alpha=-1 \rangle$ states indicated by round arrows in Fig.~\ref{fig:setup}(d) are nearly at resonance with the frequency $\omega$. 
The amplitude of the scattering between the resulting p-h pairs brings for each factor of the interaction strength the singular combination $(\omega - \delta \omega)^{-1}$.
The above scattering processes are exemplified in Fig.~\ref{fig:scattering}(a).
The other class of equally singular contributions is the self-energy corrections (see Fig.~\ref{fig:scattering}(b)).
Indeed, the singularity in the propagator of a nearly degenerate pair is obtained when both particle and hole in the p-h pair are at the mass shell.
The self-energy insertions account for the on-shell singularities of particle and hole propagators separately, and should be therefore included on equal footing with the inter-pair scattering.

\begin{figure}[ht!]
\begin{center}
\includegraphics[width=0.9\columnwidth]{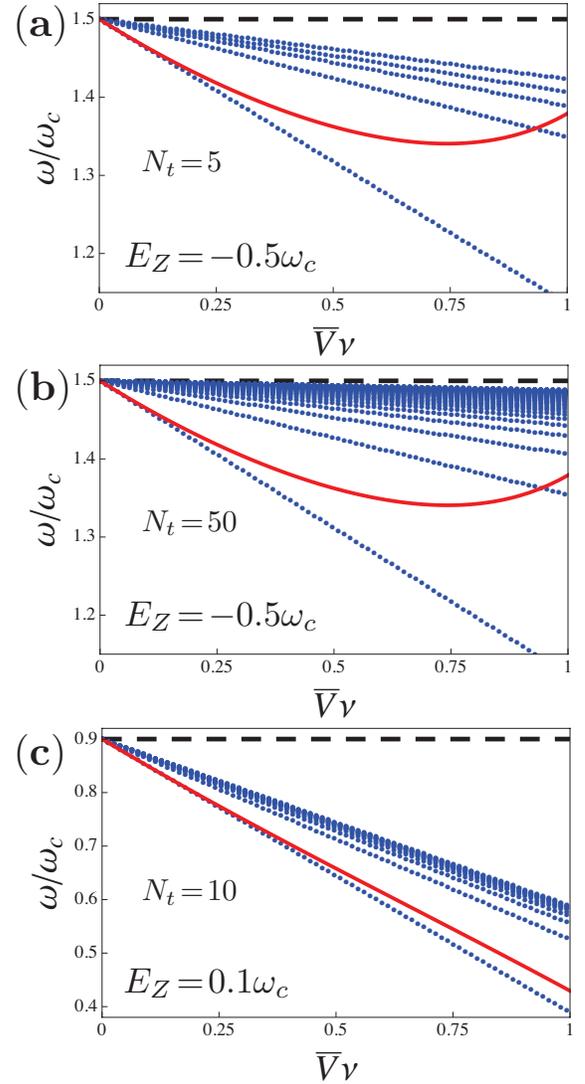}
\caption{ (color online) The dotted (blue) lines show the spectrum of collective modes that are superpositions of p-h excitations between the subbands $|n+1,+1\rangle$ and $|n,-1\rangle$ as a function of the interaction parameter $\bar{V} \nu =(V_{\rho} - 3 V_s) \nu$.
The solid (red) line is the frequency of the spin-sloshing mode as given by Eq.~\eqref{result1_a} obtained for the weakly interacting Fermi liquid. 
The horizontal dashed lines indicate the non-interacting excitation energy in units of $\omega_c$, $|\omega_c - E_Z|/\omega_c$.
The parameters used are (arb. units) $E_F = 50$,  $m=1$ and (a) $E_Z = -0.5 \omega_c$, $\omega_c = 10$, giving $N_t = 5$ according to Eq.~\eqref{Nt}; (b) $E_Z = -0.5 \omega_c$, $\omega_c = 1$, resulting in $N_t=50$ modes;
(c) $E_Z = 0.1 \omega_c$, $\omega_c = 5$, giving $N_t = 10$ modes.
The spin sloshing mode corresponds to the lowest energy mode of collective excitations.}
\label{fig:res1}
\end{center}
\end{figure}

To include most singular vertex and self energy corrections it is convenient to introduce the set of operators
\begin{align}\label{P}
P^{\dag}_{n} = \sum_{k} \psi_{n+1,+1, k}^{\dag} \psi_{n,-1, k}
\end{align}
creating the near degenerate p-h excitations.
The frequency of the collective spin-sloshing mode is given by the poles of the matrix correlation function
\begin{align}\label{Gamma}
\hat{\Gamma}_{nn'}(\omega) = \int_{-\infty}^{\infty} d t e^{  i \omega t}\left\langle T_t P_n(0) P^{\dag}_{n'}(t) \right\rangle\, ,
\end{align}
in a complex $\omega$ plane, and $T_{t}$ stands for the time ordering operation.
The indices $n$ and $n'$ running respectively over the $N_t$ rows and $N_t$ columns of the matrix $\hat{\Gamma}$ in Eq.~\eqref{Gamma} label the pairs of bands of the form $\{|n,-1\rangle, |n+1,+1 \rangle \}$ that can host p-h excitations created by the operators Eq.~\eqref{P}.
It follows that the dimensionality of $\hat{\Gamma}$, $N_t$ is equal to the number of such pairs with at least one subband crossing the Fermi level.
We have,
\begin{align}\label{Nt}
N_t = \mathrm{Int} 
\left[ \frac{E_F}{\omega_c} + \frac{1}{2} \left\vert \frac{|E_Z|}{\omega_c} - 1\right\vert \right]\, ,
\end{align}
where $\mathrm{Int} \left[ x \right]$ stands for the integer part of $x$.

The  summation of the most singular contributions amounts to solving the Dyson equation presented graphically in Fig.~\ref{fig:scattering}(c).  
This procedure amounts to the random phase approximation used to study the dispersion relation of collective modes in quasi-one-dimensional wires \cite{Li1989,Haupt1991}. 
The intersubband spin plasmons in quantum wells were studied within a density-functional formalism \cite{Ullrich2002,Ullrich2003}.

The Dyson equation is solved  in the matrix form,
\begin{align}\label{Gamma_D}
\hat{\Gamma}(\omega) = \left[ \hat{\Pi}^{-1}(\omega) + \hat{W} \right]^{-1}\, .
\end{align}
The frequencies of the collective excitations $\omega_{\mathrm{coll}}$ satisfy 
\begin{align}\label{det}
\det \left[ \hat{\Pi}^{-1}(\omega_{\mathrm{coll}}) + \hat{W} \right] = 0\, .
\end{align}
In the expression \eqref{Gamma_D} the polarization operator $\hat{\Pi}(\omega)$ is a diagonal matrix,
$\left[\hat{\Pi}\right]_{nn'}(\omega) = \delta_{nn'} \Pi_{n}(\omega)$,
\begin{align}\label{Polarization}
\Pi_{n}(\omega) = 
\int \frac{d k d \epsilon}{ (2 \pi)^2 }
G_{n+1,+1}(\epsilon+\omega,k)
G_{n,-1}(\epsilon,k)\, ,
\end{align}
where Green functions,
\begin{align}\label{Green}
G_{n\alpha}(\epsilon,k) = \left[ \epsilon - (E_{n\alpha}(k)-E_F)   - \Sigma_{n,\alpha}\right]^{-1}  
\end{align}
contains the self energy $\Sigma_{n,\alpha}$ calculated to the first order in interaction, 
\begin{align}\label{self1}
\Sigma_{m,\alpha} & =
\bar{V} \ell^{-1} \sum_{m'=0}^{N_t} \frac{k^F_{m',-\alpha}}{ \pi } M^{m,m'}_{m,m'} \, ,
\end{align}
where the charge and spin interaction amplitudes combine into a single combination 
\begin{align}\label{bar_V}
\bar{V} = V_{\rho}- 3 V_s \, ,
\end{align}
 and 
\begin{align}\label{M}
M^{n,n'}_{l,l'} = \int_{-\infty}^{\infty} d \xi \varphi_n(\xi) \varphi_{n'}(\xi) \varphi_l(\xi) \varphi_{l'}(\xi)\, . 
\end{align}
Note that the spin up(down) self energy in Eq.~\eqref{self1} is expressed via the Fermi momenta of the spin down(up) fermions as required by Pauli principle for the point-like interaction.
Consistent with the approximations made, the higher order corrections to the self energy as well as Green functions off-diagonal in the subband index are omitted.
The vertex $\hat{W}$ matrix in Eq.~\eqref{Gamma_D} is expressed through the integrals of the type \eqref{M} as 
\begin{align}\label{W}
W_{n,n'} =\bar{V} \ell^{-1} M^{n+1,n'+1}_{n,n'},
\end{align}

The substitution of Eq.~\eqref{Green} in Eq.~\eqref{Polarization} followed by the straightforward energy and momenta integrations yields
\begin{align}\label{Polarization1}
\Pi_{n}(\omega) = 
\frac{\pi^{-1} \left( k^F_{n,-1} - k^F_{n+1,+1}\right) }{
\omega -\omega_c + E_Z
-
\Sigma_{n+1,+1} + \Sigma_{n,-1}
}\, .
\end{align}
The Fermi momenta in the numerator of Eq.~\eqref{Polarization1} are given by Eq.~\eqref{kF} for free fermions as required by the consistency with our approximations.

The explicit expression \eqref{Polarization1} allows us to reformulate the  Eq.~\eqref{det} as an eigenvalue problem amenable to numerical analysis.
Using Eq.~\eqref{Polarization1} we write
\begin{align}\label{Polarization2}
\hat{\Pi} = \hat{K}^{-1} \left( \omega - \omega_c + E_Z + \hat{\Sigma} \right)\, ,
\end{align}
where we have introduced the two diagonal matrices,
\begin{align}\label{hat_K}
\left[\hat{K}\right]_{m,m'} = \frac{1}{\pi} \left( k^{F}_{m-1,-1} - k^{F}_{m,+1} \right)\delta_{m,m'}
\end{align}
and
\begin{align}\label{hat_Sigma}
\left[\hat{\Sigma}\right]_{m,m'} = \left(- \Sigma_{m,+1} + \Sigma_{m-1,-1}\right) \delta_{m,m'} \, .
\end{align}

The equation \eqref{Polarization2} allows us to rewrite the condition Eq.~\eqref{det} as
\begin{align}\label{det1}
\det \left[  \omega_{\mathrm{coll}} - \omega_c + E_Z + \hat{\Sigma}  + \hat{W} \hat{K}  \right] = 0\, .
\end{align}
The condition \eqref{det1} implies that if $\lambda_j$ are the eigenvalues of the matrix $\hat{\Sigma}  +\hat{W} \hat{K}$ the frequencies of the collective modes are given by 
\begin{align}\label{collect}
\omega_{\mathrm{coll}} = \left| \omega_c - E_Z - \lambda_j \right|\, .
\end{align}
The expression \eqref{collect} reduces the solution of Eq.~\eqref{det} to the eigenvalue problem.

Since both the self energy $\hat{\Sigma}$ and the vertex function $\hat{W}$ are proportional to $\bar{V}$ it is clear that the frequencies of eigenmodes, Eq.~\eqref{collect} are linear functions of $\bar{V}$.
This linearity is an artifact of the weak coupling approximation, $\bar{V} \nu \lesssim 1$.
The results for the solution of the Eq.~\eqref{det1} are shown as a function of the interaction strength in Fig.~\ref{fig:res1}(a,b)  for $E_Z = -0.5 \omega_c$ and in Fig.~\ref{fig:res1}(c) for $E_Z = 0.1 \omega_c$.
\subsection{Comparison of the results obtained in Fermi liquid and by perturbation theory}
\label{sec:Comparison}
Here we compare the results obtained in Secs.~\ref{sec:Collective_spin-sloshing} and \ref{sec:Microscopic}.
We limit the discussion to the weak interaction, $V_{\rho}\nu, V_s \nu \lesssim 1$ when both approaches are applicable.

To be able to use the results Eqs.~\eqref{result1} and \eqref{result2} we have to find the expressions for the Fermi liquid parameters for the microscopic Hamiltonian specified by Eqs.~\eqref{M11} and \eqref{M19}. 
To the leading order in $V_{\rho,s}$ the only non-zero Fermi liquid parameters read 
\begin{align}\label{comp13}
F_0 = - G_0 = \frac{1}{2} \bar{V} \nu\, ,
\end{align}
where the combination $\bar{V}$ is defined by Eq.~\eqref{bar_V}.
See App.~\ref{app:Fermi} for the derivation of equation \eqref{comp13}.
To the leading order in interaction the Fermi liquid parameters in Eq.~\eqref{comp13} are position independent.
In general this is not the case, and a different procedure combing the phenomenology with the measurements is required as proposed in the context of trapped Fermi gases \cite{Chien2010}.

With the approximation \eqref{comp13} the result \eqref{result1} takes the form
\begin{align}\label{result1_a}
\omega_s \approx 
E_Z\frac{4 - \bar{V}\nu}{4 - 2\bar{V} \nu}  - \sqrt{ E_Z^2 \frac{ \bar{V}^2 \nu^2}{4( 2 - \bar{V}\nu)^2} + \omega_c^2 \frac{ 2 -\bar{V}\nu}{2 + \bar{V}\nu} }\, .
\end{align}
The expression \eqref{result1_a} is shown in Fig.~\ref{fig:res1} along with the numerical solutions for poles of Eq.~\eqref{Gamma}.
The agreement between the Fermi-liquid result \eqref{result1} and the perturbation theory is naturally achieved in the regime of weak interactions, $\bar{V}\nu \lesssim 1$.
The comparison between Fig.~\ref{fig:res1}(a) and Fig.~\ref{fig:res1}(b) indicates that the two approaches are found to agree already for $5$ occupied pairs of subbands. 

As the expression Eq.~\eqref{result1_a} is valid for $\bar{V} \nu \lesssim 1$ we expand it, and obtain the result \eqref{result1_b}.
Equations  \eqref{result1_b} and \eqref{result1_a} demonstrates the absence of Kohn theorem for the spin sloshing mode explicitly.
The frequency $\omega_s$ of the spin-sloshing mode shifts due to the interactions by a finite amount $\bar{V} \nu ( E_Z/4 + \omega_c/2 )$.

\subsection{Interaction induced shift of the BSR}
\label{sec:Interaction}

The spin sloshing mode becomes soft for the special Zeeman splitting $E_Z^*$ such that $\omega_s(E_Z^*) =0$.
Solving Eq.~\eqref{result1_b} for $E_Z^*$ to the leading order in interaction we obtain 
\begin{align}\label{EZ_star}
E_Z^* \approx \omega_c( 1 - 3 \bar{V} \nu/4 )\, .
\end{align}
Equivalently, this expression follows from the expansion of the result \eqref{result2} in $\bar{V} \nu$ using Eq.~\eqref{comp13}.
The expression \eqref{EZ_star} is the value of the Zeeman splitting at which the spin sloshing mode has a zero energy.
It has been obtained within the Fermi liquid with parameters evaluated for the weak and short range interaction, Eq.~\eqref{M19}.

\begin{figure}[h]
\begin{center}
\includegraphics[width=0.9\columnwidth]{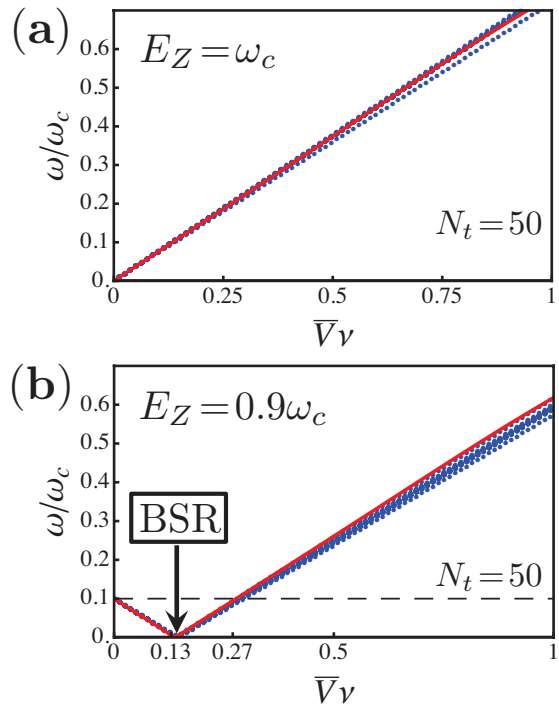}
\caption{ (color online) The dotted (blue) lines show the spectrum of p-h excitations between the subbands $|n+1,+1\rangle$ and $|n,-1\rangle$ as a function of the interaction parameter $\bar{V} \nu =(V_{\rho} - 3 V_s) \nu$.
The solid (red) line is the frequency of the spin-sloshing mode as given by Eq.~\eqref{result1_a}.
(a) At $E_Z = \omega_c$ the system is at resonance and goes off resonance at finite interaction.
(b) At $E_Z = 0.9\omega_c$ the resonance is achieved for 
$\bar{V} \nu = 40/3 \approx 1.33$ as follows from Eq.~\eqref{BSR_ren}.
The horizontal dashed lines indicate the non-interacting excitation energy  $|\omega_c - E_Z|/\omega_c$.
The parameters used are (arb. units) $E_F = 50$,  $m=1$, $\omega_c = 1$ corresponding to $N_t = 50$. 
The spin sloshing mode follows the continuum.}
\label{fig:res_comp}
\end{center}
\end{figure}

To relate Eq.~\eqref{EZ_star} to the interaction induced renormalization of the BSR we compare the spin-sloshing mode frequency given by Eq.~\eqref{result1_a} or Eq.~\eqref{result1_b} to the spectrum of the rest of the p-h excitations in the regime $\omega_c \approx E_Z$.
The results are presented in Fig.~\ref{fig:res_comp}.
 
Figure \ref{fig:res_comp} shows how the p-h excitations involving the pairs of subbands of the form $|n+1,+1\rangle$ and $|n,-1\rangle$ change there energy with interaction for $E_Z = \omega_c$.
At no interaction all such pairs are zero energy excitations which is the condition for the onset of the BSR.
It follows that even for the number of p-h excitations as large as $N_t=50$ the interaction causes the same energy shift for all the p-h excitations.
In other words, the p-h pairs stay degenerate even in the presence of interaction.
The spin sloshing mode has the same energy as the individual p-h pairs.
The reason for this is the vanishing of the polarization operator, Eq.~\eqref{Polarization1} for $k^F_{n,-1} - k^F_{n+1,+1}$.
As is evident from Fig.~\ref{fig:res_comp} that all the solutions follow the spin sloshing mode given by the Fermi-liquid expression \eqref{result1_a}. 
The spin sloshing mode is strongly coupled to p-h continuum and is a short-lived excitation.

Importantly the expression \eqref{result1_a} obtained within the Fermi liquid for its energy is at the same time the energy of the p-h excitations between the subband $|n+1,+1\rangle$ and subband $|n,-1\rangle$.
It follows that the condition for the BSR is the vanishing of $\omega_c$ 
which occurs at $E_Z = E_Z^*$ with $E_Z^*$ given by Eq.~\eqref{EZ_star}.
Then, the BSR renormalization due to interaction takes the form of Eq.~\eqref{BSR_ren}.
The result \eqref{BSR_ren} is illustrated in Fig.~\ref{fig:res_comp} where we set $E_Z = 0.9 \omega_c$ and the non-interacting system is off resonance.
Yet due to the renormalization Eq.~\eqref{BSR_ren} the system is at resonance for $\bar{V} \nu = 40/3 \approx 1.33$ as the relevant continuum of p-h pairs becomes soft.

\section{Conclusions}
\label{sec:Conclusions}

To conclude, we studied the effect of electron interaction on the collective and p-h excitations in a laterally confined two-dimensional Fermi gas.
The spectrum of confined electrons splits into subbands of transverse quantization.
We focused on the transversal excitations across the channel controlled by the subband separation energy scale.

When the Zeeman splitting is tuned to the subband separation a special type of  resonance, the BSR is detected in the dc transport measurements of Ref.~\cite{Frolov2009,Frolov2009a}.
Hence we analyzed the class of collective modes adiabatically connected to the spin flip intersubband p-h excitations which become soft at the BSR. 
The BSR measurements of \cite{Frolov2009} probe the individual p-h excitations.
We showed however relying on the perturbation theory that for a weak, short range interaction and sufficiently close to the BSR such p-h excitations are indistinguishable from the collective spin sloshing mode.
Therefore finding the frequency of the collective spin sloshing mode at the same time allows us to determine the shift of the BSR caused by interactions.

To meet this goal we applied the Fermi liquid theory to identify and analyze the spin sloshing mode.
It combines the features of two other more familiar collective excitations.
The first is the density oscillations across the channel, with amplitude vanishing at the center of the channel and is referred to as sloshing, or Kohn mode. 
The second is the collective spin precession mode in the presence of a Zeeman splitting.

The spin-sloshing mode is the collective spin precession with an amplitude growing linearly away from the center of the channel, see Fig.~\ref{fig:slosh}.
We demonstrated that the Kohn theorem does not apply to it, and found its frequency renormalization due to a weak and short range interaction.

Finally, by combining the above results we obtained the physical picture of interaction effect on the BSR.
We start by setting the Zeeman splitting to the subband separation in a Fermi gas.
Then the spin-flip intersubband p-h excitations have a zero energy, and the system is at resonance.
When a weak interaction is turned on these p-h excitations remain degenerate with their common energy becoming non-zero.
Hence, the interaction detunes the resonance.
To tune it back the Zeeman splitting must be accordingly adjusted.
It follows that the resonant magnetic field is modified by interactions. 
To quantify this statement we used the analytical expression for the energy of a spin sloshing mode was obtained within the Fermi liquid theory.

In our analysis we assumed the confining potential to be parabolic with the equidistantly separated subbands, see Eq.~\eqref{M17}.
Although the realistic confining potential is never strictly harmonic, the deviation from parabolic potential profile is in many cases small. 
If we take the non-parabolic part of the confining potential in the form,
$\delta V_{np}(z) = (\epsilon/4)z^4$ then the continuum of the p-h excitations is split due to the variation of the level shift $\delta E_{n\alpha}(k_x) \approx \langle n | \delta V_{np}| n \rangle $ with the subband index $n$.
The unharmonicity effect is negligible provided the energy shift due to the interactions, $\omega_c (V_{\rho} \! -\!  3 V_s)\nu$, Eq.~\eqref{result1_b} exceeds the typical value of $\delta E_{n\alpha}(k_x)$. 
This gives the upper bound, $\epsilon < (m \omega_c^2/ 2 E_F)^2 \omega_c m (V_{\rho} \! -\!  3 V_s)$, as we estimate $\langle z^4 \rangle \sim  (2 E_F/ m \omega_c^2)^2$.

In summary, the interaction induced shift of the BSR is obtained by tracing the frequency renormalization of the collective spin sloshing mode.
The shift of the BSR is a consequence of the non-existence of the Kohn theorem for this mode.

\acknowledgements 
We are grateful to D. H. Berman, J. C. Euges, A. M. Finkel'stein, M. E. Flatt\'e, J. A. Folk and M. E. Raikh for valuable discussions.
Authors acknowledge the support of the University of Iowa.

\begin{appendix}

\section{Fermi liquid amplitudes to the first order in interaction}
\label{app:Fermi}
\begin{figure}[h]
\begin{center}
\includegraphics[width=0.5\columnwidth]{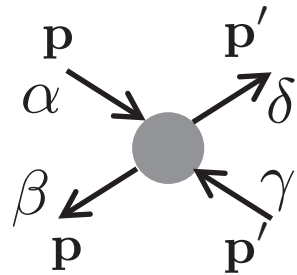}
\caption{The definition of quasi-particle momenta and spin indices of the scattering amplitude, Eq.~\ref{g_omega}}
\label{fig:app_ampl}
\end{center}
\end{figure}

Consider the interaction of the form Eq.~\eqref{M19}.
As this interaction is point-like the scattering amplitudes have no angular dependence,
\begin{align}\label{g_omega}
\Gamma^{\omega}_{\vec{p}\vec{p}'} = &  V_{\rho} \left[  \delta_{\alpha \beta} \delta_{\gamma \delta} - \delta_{\alpha\gamma} \delta_{\beta\delta} \right]
\notag \\
& +
V_s \left[  \vec{\sigma}_{\alpha\beta} \vec{\sigma}_{\gamma\delta} 
- \vec{\sigma}_{\alpha\delta} \vec{\sigma}_{\gamma\beta} \right]\, ,
\end{align}
where the superscript $\omega$ is introduced as the ladder diagrams we considered in Sec.~\ref{sec:Microscopic} are taken in the limit of zero total momentum of p-h pairs.
The momenta and spin indicies used in Eq.~\eqref{g_omega} are defined in Fig.~\ref{fig:app_ampl}. 
Both density and spin interaction channels contribute to the scattering amplitude via the direct and exchange interaction processes as shown in Fig.~\ref{fig:scattering}(a).

The amplitude in Eq.~\eqref{g_omega} is directly related to the functions $ f_{\vec{p}\vec{p}'}$ and $ g_{\vec{p}\vec{p}'}$ entering the phenomenological relation \eqref{SP1}, \cite{Pitaevskii1980}
\begin{align}\label{g_omega1}
\nu \Gamma^{\omega}_{\vec{p}\vec{p}'} = f_{\vec{p}\vec{p}'} \delta_{\alpha \beta} \delta_{\gamma \delta} 
+
g_{\vec{p}\vec{p}'} \vec{\sigma}_{\alpha\beta} \vec{\sigma}_{\gamma\delta}\, . 
\end{align}
For short range interaction, and to the first order in interaction $f_{\vec{p}\vec{p}'} = F_0$ and $g_{\vec{p}\vec{p}'} = G_0$.
Furthermore using the identity,
\begin{align}\label{identity}
\delta_{\alpha\delta} \delta_{\beta \gamma} = \frac{1}{2}\vec{\sigma}_{\alpha\beta} \vec{\sigma}_{\gamma\delta} +\frac{1}{2} \delta_{\alpha\beta} \delta_{\gamma\delta}
\end{align}
to rewrite Eq.~\eqref{g_omega} in the form of Eq.~\eqref{g_omega1} we arrive at Eq.~\eqref{comp13}  of the main text.

\section{Sloshing and spin precession modes: perturbation theory analysis}
\label{app:Kohn}

\begin{figure}[h]
\begin{center}
\includegraphics[width=0.7\columnwidth]{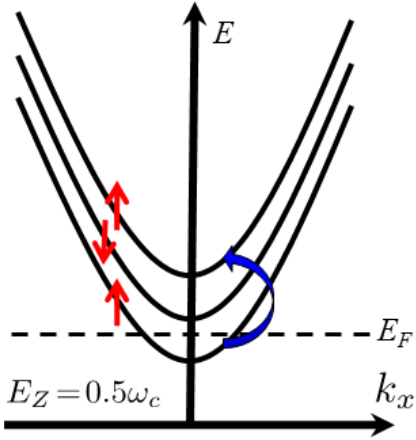}
\caption{(color online) The three lowest energy subbands are $|0,+1\rangle$, $|0,-1\rangle$ and $|1,+1\rangle$.
The spin polarization is denoted by straight thin (red) arrows. 
Only the $|0,+1\rangle$ crosses the Fermi level, $E_F$.
The sloshing mode is the superposition of  p-h excitations from $|0,+1\rangle$ to $|1,+1\rangle$ subbands as indicated by a round thick (blue) arrow.}
\label{fig:par1}
\end{center}
\end{figure}

\begin{figure}[h]
\begin{center}
\includegraphics[width=0.9\columnwidth]{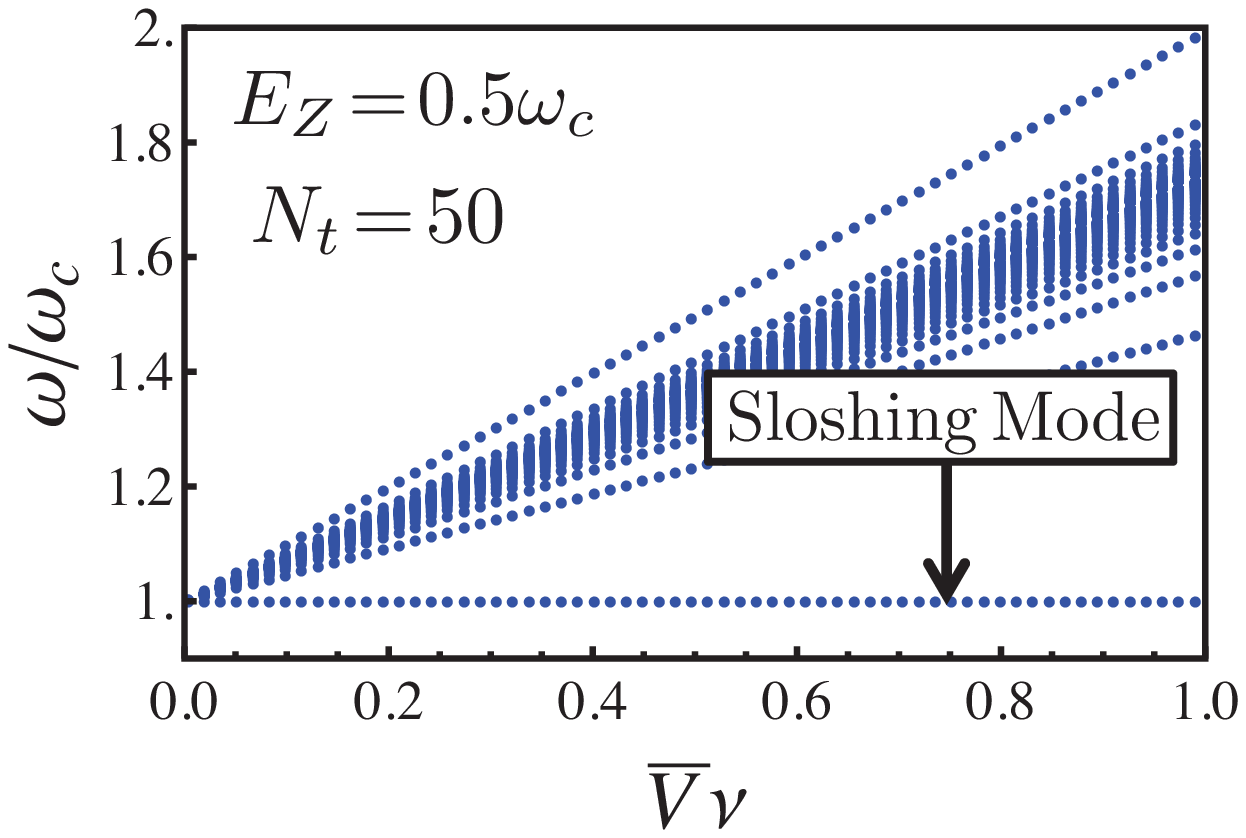}
\caption{(color online) The dotted (blue) lines show the spectrum of collective modes that are the superpositions of p-h excitations between the subbands $|n+1,\pm1\rangle$ and $|n,\pm1\rangle$ as a function of the interaction parameter $\bar{V} \nu =(V_{\rho} - 3 V_s) \nu$.
The parameters used are (arb. units) $E_F = 50$,  $m=1$, $E_Z = 0.5 \omega_c$ and $\omega_c = 1$ corresponding to $N_t = 50$. 
The horizontal dotted line at $\omega/\omega_c =1$ demonstrates the Kohn theorem for sloshing mode.}
\label{fig:app_slosh}
\end{center}
\end{figure}

Here we demonstrate the consistency of the perturbation theory with the variants of Kohn theorem for the sloshing mode and spin precession mode.
We stress that although the perturbation theory holds for weak interaction the cancellation between the self-energy and vertex corrections holds exactly in every order of perturbation theory.
Such cancellation provides us with a test of the numerical calculations.

\subsection{sloshing mode}
Here we demonstrate the Kohn theorem for the sloshing mode in the presence of Zeeman splitting and electron-electron interaction, Eq.~\eqref{M19}.
The relevant p-h excitations are formed by the spin conserving transition between the adjacent subbands.
The electron and hole comprising such a pair reside at subbands $|n,\pm1\rangle$ and $|n+1,\pm 1\rangle$ respectively.

The most trivial is the situation with only the very lowest subband $|0,+1\rangle$ occupied, see Fig.~\ref{fig:par1}.
As only spin up species are present the short range interaction is ineffective for fermions, and the Kohn theorem is trivially satisfied.
Technically the contributions from the direct and exchange processes to both the self energy and the vertex corrections cancel.
The situation is less trivial for the large number of occupied levels.
In all cases the sloshing mode can be identified as the one not renormalized by interactions, see Fig.~\ref{fig:app_slosh}.

\begin{figure}[h]
\begin{center}
\includegraphics[width=0.7\columnwidth]{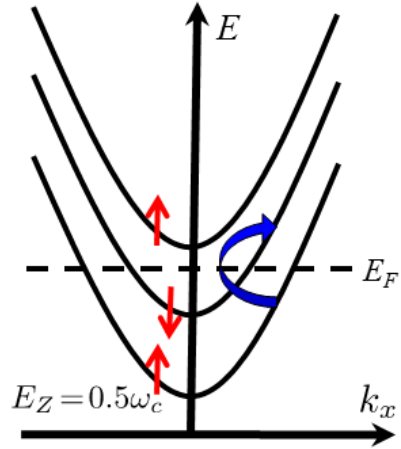}
\caption{(color online) The same three lowest energy subbands, $|0,+1\rangle$, $|0,-1\rangle$ and $|1,+1\rangle$ as in Fig.~\ref{fig:par1}.
The spin polarization is denoted by straight thin (red) arrows. 
The subbands $|0,+1\rangle$ and $|0,-1\rangle$ cross the Fermi level, $E_F$.
The spin precession mode is the superposition of  p-h excitations from $|0,+1\rangle$ to $|0,-1\rangle$ subbands as indicated by a round thick (blue) arrow.}
\label{fig:par2}
\end{center}
\end{figure}

\begin{figure}[t!]
\begin{center}
\includegraphics[width=0.9\columnwidth]{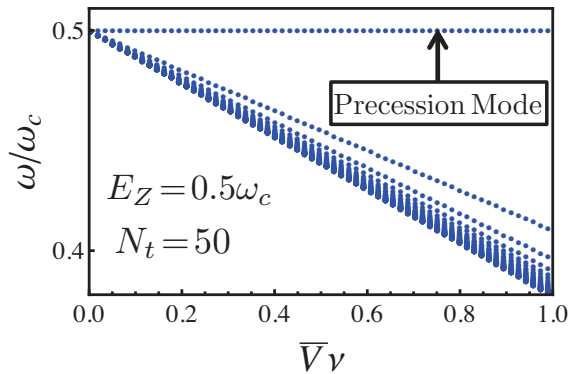}
\caption{ (color online) The dotted (blue) lines show the spectrum of collective modes that are the superpositions of p-h excitations between the subbands $|n,+1\rangle$ and $|n,-1\rangle$ as a function of the interaction parameter $\bar{V} \nu =(V_{\rho} - 3 V_s) \nu$.
The parameters used are (arb. units) $E_F = 50$,  $m=1$, $E_Z = 0.5 \omega_c$ and $\omega_c = 1$ corresponding to $N_t = 50$. 
The horizontal dotted line at $\omega/ \omega_c=(\omega_c - E_Z)/\omega_c = 0.5 \omega_c$ demonstrates the Kohn theorem for spin precession mode.}
\label{fig:app_precession}
\end{center}
\end{figure}

\subsection{spin precession mode}
\label{sec:app_precession}
Here we make another generalization of the procedure developed in the Sec.~\ref{sec:Microscopic} to describe the effect of interactions on intra-band spin-flip transitions.
We start with the illustration of the Kohn theorem for the simplest case of two subbands occupied with Fermi momenta $k^F_{0,-1}$ and $k^F_{0,+1}$, see Fig.~\ref{fig:par2}. 
It is sufficient to focus on a single type of p-h excitations from subband $|0,+1 \rangle$ to subband $|0,-1 \rangle$.

In this case the proper generalization of the $\hat{\Gamma}$ matrix introduced in Eq.~\eqref{Gamma} has a single element,
\begin{align}\label{Gamma_app}
\Gamma(\omega) =\left[ \Pi^{-1}(\omega) + W \right]^{-1}\, ,
\end{align}
where we have for the polarization operator describing the excitations of p-h pairs shown in Fig.~\ref{fig:par2}, 
\begin{align}\label{Polarization1_app}
\Pi(\omega)= 
\frac{\pi^{-1} \left( k^F_{0,+1} - k^F_{0,-1}\right) }{
\omega  - E_Z
-
\Sigma_{0,-1} + \Sigma_{0,+1}
}
\end{align}
instead of Eq.~\eqref{Polarization1}.
The Fermi momenta and the self energies entering Eq.~\eqref{Polarization1_app} are defined by Eqs.~\eqref{kF} and \eqref{self1} respectively.
Instead of Eq.~\eqref{W} we have for the scattering vertex,
\begin{align}\label{W_app}
W =\bar{V} \ell^{-1} M^{0,0}_{0,0} \, .
\end{align}
It follow from Eq.~\eqref{Polarization1_app} that
\begin{align}\label{Polarization2_app}
\Pi^{-1}(\omega=E_Z)= 
\frac{\pi (\Sigma_{0,+1}-\Sigma_{0,-1})}{  k^F_{0,+1} - k^F_{0,-1}}\, .
\end{align}
Equation \eqref{self1} gives
\begin{align}\label{self1_app}
\Sigma_{0,\pm 1} & =
\bar{V} \ell^{-1}  \frac{k^F_{0,\mp 1}}{ \pi } M^{0,0}_{0,0}\, .
\end{align}
Combining Eqs.~\eqref{Gamma_app}, \eqref{W_app}, \eqref{Polarization2_app}, and \eqref{self1_app} we obtain
\begin{align}
\Gamma^{-1}(\omega = E_Z) = 0
\end{align}
which signifies the presence of the collective mode with unrenormalized frequency, $E_Z$ as expected.

The outlined derivation is easily generalized to arbitrary number of occupied subbands.
The condition analogous to Eq.~\eqref{det} leads to the spectrum of excitation shown in Fig.~\ref{fig:app_precession}.
The spin precession mode is identifiable as the one not renormalized by interactions.
Essentially similar approach has been used for the calculation of the spin wave dispersion in quantizing magnetic field with unequal occupation of Zeeman split Landau levels, (see Ref.~\cite{Kallin1984}).

\end{appendix}

\end{document}